\documentclass[manuscript]{aastex63}

\submitjournal{ApJS}

\shorttitle{GCR variations in the recent solar cycles}
\shortauthors{Fu et al.}
\graphicspath{{./}}
\begin{document}

\title{Variations of the Galactic Cosmic Rays in the Recent Solar Cycles}

\correspondingauthor{Xiaoping Zhang}
\email{xpzhangnju@gmail.com}

\correspondingauthor{Lingling Zhao}
\email{lz0009@uah.edu}

\author[0000-0003-4245-3107]{Shuai Fu}
\affiliation{State Key Laboratory of Lunar and Planetary Sciences, Macau University of Science and Technology, Taipa 999078, Macau, PR China}
\affiliation{CNSA Macau Center for Space Exploration and Science, Taipa 999078, Macau, PR China}

\author[0000-0002-4306-5213]{Xiaoping Zhang}
\affiliation{State Key Laboratory of Lunar and Planetary Sciences, Macau University of Science and Technology, Taipa 999078, Macau, PR China}
\affiliation{CNSA Macau Center for Space Exploration and Science, Taipa 999078, Macau, PR China}

\author[0000-0002-4299-0490]{Lingling Zhao}
\affiliation{Center for Space Plasma and Aeronomic Research (CSPAR), University of Alabama in Huntsville, Huntsville, AL 35805, USA}

\author[0000-0002-5565-8382]{Yong Li}
\affiliation{State Key Laboratory of Lunar and Planetary Sciences, Macau University of Science and Technology, Taipa 999078, Macau, PR China}
\affiliation{CNSA Macau Center for Space Exploration and Science, Taipa 999078, Macau, PR China}

\begin{abstract}

In this paper, we study the galactic cosmic ray (GCR) variations over the solar cycles 23 and 24, with measurements from the NASA's \emph{ACE}/CRIS instrument and the ground-based neutron monitors (NMs). The results show that the maximum GCR intensities of heavy nuclei ($5 \leqslant Z \leqslant 28$,  50$\sim$500 MeV/nuc) at 1 AU during the solar minimum in 2019--2020 break their previous records, exceeding those recorded in 1997 and 2009 by $\sim$25\% and $\sim$6\%, respectively, and are at the highest levels since the space age. However, the peak NM count rates are lower than those in late 2009. The difference between GCR intensities and NM count rates still remains to be explained. Furthermore, we find that the GCR modulation environment during the solar minimum $P_{24/25}$ are significantly different from previous solar minima in several aspects, including remarkably low sunspot numbers, extremely low inclination of the heliospheric current sheet, rare coronal mass ejections, weak interplanetary magnetic field and turbulence. These changes are conducive to reduce the level of solar modulation, providing a plausible explanation for the record-breaking GCR intensities in interplanetary space.

\end{abstract}

\section{Introduction} \label{sec:intro}
Due to the supersonic expansion of the solar wind (SW), the heliosphere is like a giant asymmetric bubble filled with various energetic particles. These particles are the dominant source of space radiation environment, posing a significant threat to the regular operation of spacecraft and high-altitude aircraft, and the health of astronauts and flight personnel (e.g., \citealt{Mertens19}). Solar energetic particles (SEPs), galactic cosmic rays (GCRs) and  anomalous cosmic rays (ACRs) are the three commonly seen energetic particles in the interplanetary space. 

SEPs, also known as solar cosmic rays, are the particles originating from the Sun with energies ranging from a few tens of keV to several GeV. They are frequently associated with coronal mass ejections (CMEs) and solar flares. It is widely accepted that CME-driven shock is an essential site for accelerating solar energetic particles (e.g., \citealt{hu17,hu18,fu19,ding20}). GCRs are often regarded as a stable background, and are believed to originate from the supernova remnants (SNRs) and accelerated by supernova blast shocks driven by expanding SNRs. This is confirmed by observations and numerical simulations \citep{Aharonian07,Aharonian11,Ptuskin10}. During the quiet periods, the fluxes of some cosmic ray species with high first ionization potentials, including H, He, C, N, O, Ne and Ar, are significantly higher than the background GCRs, i.e., anomalous cosmic rays (\citealp{Hovestadt73,McDonald74,Hasebe97}).
A classical explanation of ACRs is that they originate from the interstellar medium, in which the neutral atoms drift into the heliosphere and become singly ionized by the solar wind or the ultraviolet radiation which are subsequently accelerated to energies above 10 MeV/nuc, mostly occurring at or near the heliospheric termination shock \citep{mewaldt93,Cummings07,Gloeckler09}. New mechanisms for ACR acceleration have been put forward (e.g., \citealp{Giacalone12}, and references therein), such as, \citealp{zhao19} suggested that ACR protons are accelerated in the heliosheath via reconnection processes. Studying the nature of energetic particles provides insight into the acceleration mechanisms and the transport processes between the sources and the observers. 

When GCR particles from the outer space (the so-called primary cosmic ray) encounter the Earth and penetrate the Earth's atmosphere, they interact with atmospheric atoms and generate cascades of secondary particles (e.g., \citealp{mishev14}). The secondary particles (predominantly neutrons and muons at the ground level) can be measured by the ground-based neutron monitors (NMs) and muon detectors once arriving at the surface of the Earth. The history of neutron monitors can be traced back to the 1950s \citep{bieber13}. The combination of GCR intensities from in-situ spacecrafts and GCR count rates from NM stations is helpful to sketch a relatively complete picture of cosmic rays (CRs). In addition, the cosmogenic isotope recorded in tree rings and ice cores can extend our understanding of GCRs over a longer time scale (e.g., \citealp{owens13}). 

The transport of cosmic rays throughout the heliosphere is significantly influenced by the large-scale solar wind flow and the turbulent interplanetary magnetic field embedded in it, collectively referred to as ``solar modulation". Four major modulation processes have been well described by the Parker transport equation \citep{parker65}, including gradient and curvature drifts, diffusion through the irregular interplanetary magnetic field (IMF), convection in the radial expanding solar wind, and adiabatic deceleration (or adiabatic energy loss) (e.g., \citealp{Sabbah00,zhao15,Ihongo16,zhao18}). Secular measurements provide direct evidence for evaluating their relative role in affecting cosmic ray propagation over the last few decades \citep{kota13}, but the numerical simulation of cosmic ray transport is still a highly challenging task. The solar magnetic polarity (conventionally described as [$qA$], where [$q$] denotes a positively-charged particle) is an indispensable factor in modulating cosmic ray intensities. Commonly, the positively charged particles drift inward along the heliospheric current sheet (HCS) when the heliospheric magnetic field at the north pole points inward (negative polarity, $qA<0$) and the charged particles drift outward along the HCS when the heliospheric magnetic field points outward at the north pole (positive polarity, $qA>0$) (e.g., \citealp{jokipii81,Belov00,Thomas14}). The 22-year heliomagnetic cycle (also called Hale cycle) results in the alternating sharp (negative polarity, $qA<0$) and flat-topped (positive polarity, $qA>0$) shape of cosmic ray intensity \citep{McDonald10}. 

In late 2009, both the GCR intensities measured from near-Earth spacecrafts and the GCR count rates measured from ground-based NM stations reached their then all-time maximum levels (e.g., \citealt{McDonald10,leske13}). At that time, the heliospheric environment was unusual in several aspects, including weakened interplanetary magnetic field and turbulence level, reduced solar wind dynamic pressure, and prolonged solar minimum \citep{mewaldt10}. The small magnitude of the IMF causes higher particle drift velocities, and the weakened IMF turbulence leads to larger particle mean free paths (MFPs) in the solar minimum 2009 than those in its previous solar minima. These anomalous interplanetary behaviours would reduce the level of GCR modulation and contribute to the unexpected high GCR intensities in 2009. As reported, the solar cycle 24 was the weakest in magnitude in the space age \citep{hajra21}, and it also exhibited some abnormal signs, such as, decreased sunspots (especially large spots) \citep{chapman14}, reduced coronal mass ejection events \citep{wang14}, and very flattened HCS. The solar activity seems to be going through a period of “grand minimum” \citep{jiang18,upton18,goncalves20}. Former studies have predicted that the peak of the GCR intensities in the solar minimum 24/25 (hereafter solar minimum $P_{24/25}$) would exceed or close to the level recorded in 2009--2010 (e.g., \citealp{Strauss2014, kuznetsov17,fu20}).

The main motivation of this paper is to provide a close re-examination of the GCR variations during the successive solar cycles 23 and 24, with particular interest to the maximum GCR intensities observed at 1 AU during the solar minimum $P_{24/25}$ epoch. The paper is constructed as follows. In Section \ref{sect2}, we describe the sources of the dataset utilized in this work. In Section \ref{sect3}, we present the observational results of GCRs and investigate the possible reasons for the record-breaking GCR intensities in 2019--2020 observed at 1 AU. We summarize this work in Section \ref{sect4}.

\section{Data Description}
\label{sect2}
Solar energetic particles and anomalous cosmic rays do not contribute to the intensities measured by \emph{ACE}/CRIS, and hence they do not participate in the following analysis.

GCR intensities at 1 AU are observed by the in-situ Cosmic Ray Isotope Spectrometer (CRIS) instrument on board the NASA's \emph{Advanced Composition Explorer} (\emph{ACE}) spacecraft, and GCR count rates at the ground level are monitored by the five neutron monitor stations listed in Table \ref{table1}. 

The \emph{ACE} spacecraft was launched on 25 August 1997 and has been continuously monitoring solar wind plasma, interplanetary magnetic fields and energetic particles (including both solar energetic particles and cosmic rays) at the Sun-Earth $L$1 Lagrange point \citep{stone98} for nearly 24 years, spanning the solar cycles 23 and 24. The CRIS instrument was designed to measure GCR intensities of 24 heavy species (from boron to nickel, within the energy range 50--500 MeV/nuc). Despite the low abundance ($\sim$1\%), the high atomic number and energy nuclei carry abundant information of the origin of the cosmic rays, and are of special significance to space radiation (e.g., \citealp{zhao13,fu20}). With the large geometric acceptance and high charge and mass resolution, the CRIS instrument records the most detailed and statistically significant GCR data to date \citep{stone98}. Historical \emph{ACE}/CRIS observations are well documented and publicly accessible from the ACE Science Center (ASC) via \url{http://www.srl.caltech.edu/ACE/ASC/index.html}. Here we apply the re-evaluated level-2 CRIS products, and these data are organized into 27-day Bartels rotation averages for the period from 25 August 1997 through 31 October 2020.

Pressure-corrected and 27-day averaged GCR count rates are derived from five neutron monitor stations (HRMS, JUNG, NEWK, OULU, THUL) over the 1968--2020 time period. The detailed information of the five NM stations is given in Table \ref{table1}, where $P_c$ is the local geomagnetic cutoff rigidity of each NM station, the characteristic energy (i.e., the median energy) $E_M$ is defined so that cosmic rays with energy higher (or lower) than $E_M$ contribute to half of the detector’s counting rate \citep{Usoskin08,zhao16}. All data are acquired from the Neutron Monitor Data Base (NMDB, \url{http://www01.nmdb.eu/}).

The monthly mean sunspot number (SSN) is obtained from the Solar Influences Data Analysis Center (WDC-SILSO, \url{http://www.sidc.be/}). The SW speed ($V_{sw}$), the SW dynamic pressure ($P_{d}$), the IMF magnitude ($B$) and the root mean square variation in the vector IMF ($\delta B$) are acquired from the OMNIWeb databasde (\url{https://omniweb.gsfc.nasa.gov/ow.html}). The HCS tilt angle and the mean solar polar magnetic field strength are obtained from the Wilcox Solar Observatory database (\url{http://wso.stanford.edu/}). The mean polar field is defined as $(N-S)/2$, where $N$ is the polar field strength at the northern pole, $S$ is the polar field strength at the southern pole. The halo CME lists are collected from the $SOHO$/LASCO HALO CME catalog (\url{https://cdaw.gsfc.nasa.gov/CME_list/halo/halo.html}), and the annual CME rates are from CACTus CME catalog (\url{http://www.sidc.be/cactus/}). Besides, the GCR radiation dose rates near the lunar surface are measured by the Cosmic Ray Telescope for the Effects of Radiation (CRaTER) instrument on board the \emph{Lunar Reconnaissance Orbiter} (\emph{LRO}) and are available at \url{https://crater-web.sr.unh.edu/}. Note that unless explicitly stated otherwise, all of the above-mentioned data are processed as 27-day average.

\begin{deluxetable}{cccCCCC}[htbp]
\tablecaption{List of the neutron monitor stations used in this work}
\tablehead{
\colhead{NM Station} & \colhead{Abbrev.} & \colhead{Longitude (deg)} & \colhead{Latitude (deg)} & \colhead{Altitude (m) } & \colhead{$P_c$ (GV)} & \colhead{$E_M$ (GeV)}}
\startdata
    Hermanus & HRMS  & 19.22 & -34.42 & 26    & 4.58  & 12.67 \\
    Jungfraujoch & JUNG  & 7.98  & 46.55 & 3570  & 4.49  & 12.58 \\
    Newark & NEWK  & -75.75 & 39.68 & 50    & 2.40   & 10.99 \\
    Oulu  & OULU  & 25.47 & 65.05 & 15    & 0.81  & 10.30 \\
    Thule & THUL  & -68.7 & 76.5  & 26    & 0.30   & 10.17 \\
\enddata
\tablecomments{$P_c$ is the NM's local geomagnetic cutoff rigidity, $E_M$ is the median energy of each NM station and is defined as: $E_M = 0.0877\cdot P_c^2 + 0.154\cdot P_c + 10.12$ \citep{jamsen07,Usoskin08}. }
\label{table1}
\end{deluxetable}

\section{Results and discussions}
\label{sect3}

\begin{figure}[ht!]
\epsscale{0.75}
\plotone{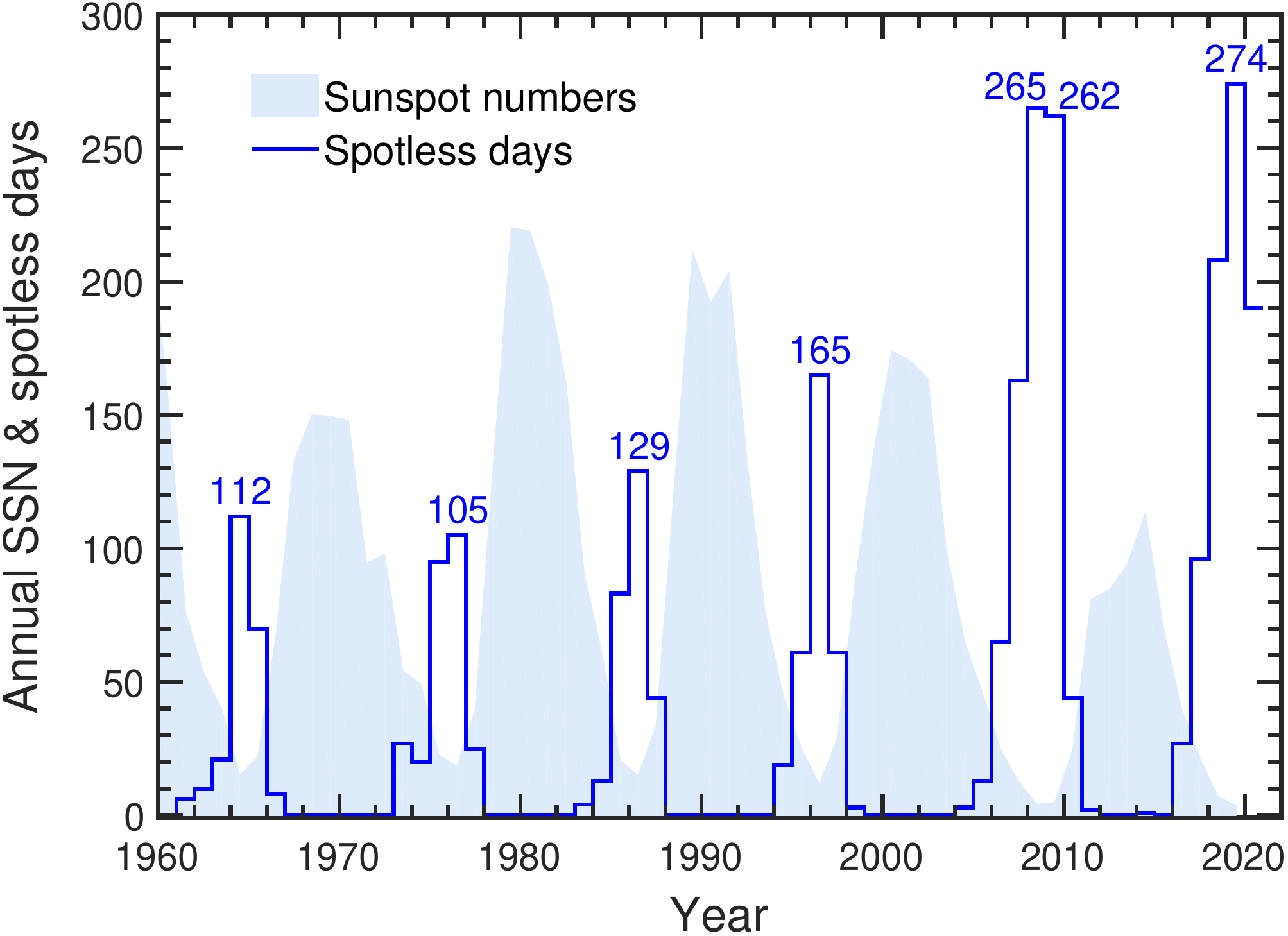}
   \caption{Yearly averaged sunspot numbers (shaded) and annual spotless days (histogram) (until October 31, 2020).
   \label{Fig1}}
\end{figure}

The SSN observations reveal that the solar activity has been persistently declining since the early 1980s. As shown in Figures \ref{Fig1} and \ref{Fig2}(a), the maximum smoothed SSNs of solar cycles 21 through 24 are 232.9, 212.5, 180.3, and 116.4, respectively, showing a prominent reduction in the solar cycle amplitude. GCR transport in the heliosphere is sensitive to varying solar activity. An intense solar activity (such as CME and solar flare) can trigger a series of space weather effects and effectively prevent GCR particles from entering the solar system (strong solar modulation), while a quiet Sun is conductive to enhance GCR fluxes as a consequence of the weakened modulation level. Next, we review the variations of the solar wind and interplanetary parameters from solar cycles 20 to 24 in Subsection \ref{subsect3.1}; GCR observational results, including GCR intensities in interplanetary space and GCR counts at the ground level, are displayed in Subsection \ref{subsect3.2}; in Subsection \ref{subsect3.3}, we investigate how the inner heliospheric environment influences the GCR intensity, with an emphasis on the record-breaking fluxes during the solar minimum $P_{24/25}$; in Subsection \ref{subsect3.4}, we point out a sudden dip in the GCR intensity during the descending phase of the solar cycle 24 and give a brief analysis of the possible causes; the measured galactic cosmic radiation doses on the lunar surface are exhibited in Subsection \ref{subsect3.5}.

\subsection{A glance of Solar Cycles 20--24}
\label{subsect3.1}

Sunspots were first observed by a telescope in the early 1600s, and the continuous daily observations began at the Zurich Observatory in the year 1849. The number of sunspots varies periodically with time and exhibits a remarkable quasi-11 year cycle, which is now widely recognized as a representative of solar activity. In Figure \ref{Fig1}, a striking feature is that the maximum amplitude of solar cycle has been decreasing since the solar cycle 21, and the recently complete solar cycle 24 is recorded to be the weakest one in the era of human space exploration \citep{hajra21}. Besides, there are totally 274  days without sunspots in the year 2019, which is the most in the past 107 years (since the year 1914). The extraordinarily quiet solar minimum offers us an opportunity to go deep into the conditions prevailing in the past grand minima periods, such as the Maunder Minimum (1645--1715) and the Dalton Minimum (1790--1830). 

\begin{figure}[ht!]
\epsscale{1}
\plotone{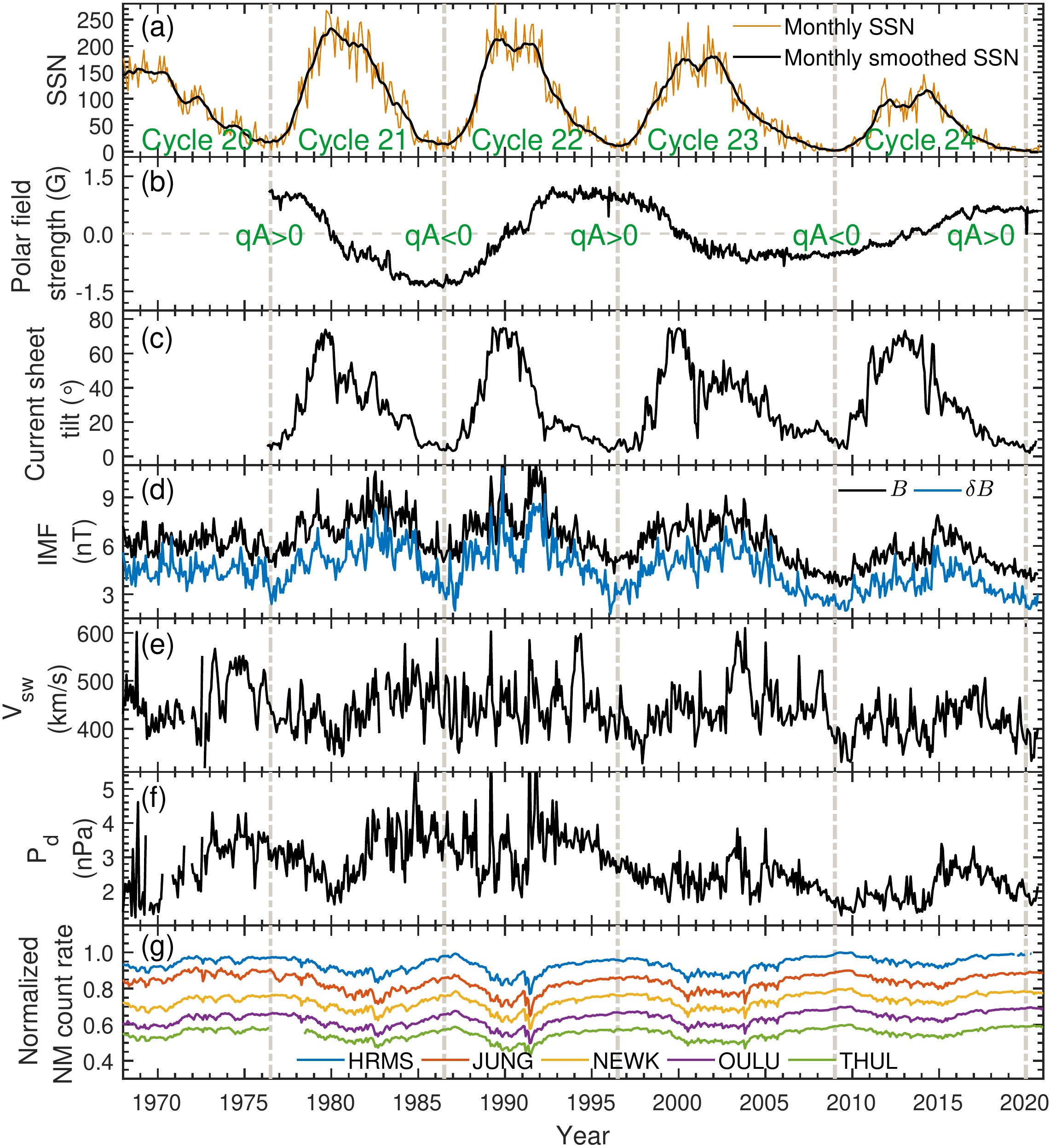}
\caption{27-day average solar wind/interplanetary parameters from January 1, 1968 to October 31, 2020. (a) Sunspot number. (b) Mean solar polar field strength. (c) HCS Tilt angle. (d) IMF magnitude $B$ and its root mean square $\delta B$. (e) Solar wind speed $V_{sw}$. (f) Solar wind dynamic pressure $P_d$. (g) Pressure-corrected neutron monitor count rates (normalized to the solar minimum $P_{23/24}$ and multiplied by arbitrary factors for distinguishing different stations). The shown magnetic field polarity in panel (b) corresponds to that at the northern pole. The vertical dashed line denotes the epoch of solar minimum.  \label{Fig2}}
\end{figure}

In Figure \ref{Fig2}, we present the variations of the solar wind/interplanetary parameters in panels (b)--(f) and the measured NM count rates in panel (g). Some well-known features of the heliosphere and GCRs can be seen from these panels. (1) Both the HCS tilt angle, the IMF strength $B$ and its turbulence $\delta B$ (i.e., root mean square of vector IMF) are positively correlated with SSN; (2) the GCR count rates at the ground level are inversely related to SSN; (3) the solar magnetic field reverses its polarity every 11 years, namely, 22-year heliomagnetic cycle (or Hale cycle); (4) the alternative peaked (negative polarity, $qA<0$) and plateau (positive polarity, $qA>0$) shapes of NM count rates during successive solar minima periods is in essence a manifestation of the Sun's polar magnetic field reversal, and the maximum NM count rate during $qA>0$ cycles (1977, 1997, 2019) is approximately 3\% lower than that during $qA<0$ cycles (1987, 2009). 

\begin{figure}[ht!]
\epsscale{1.0}
\plotone{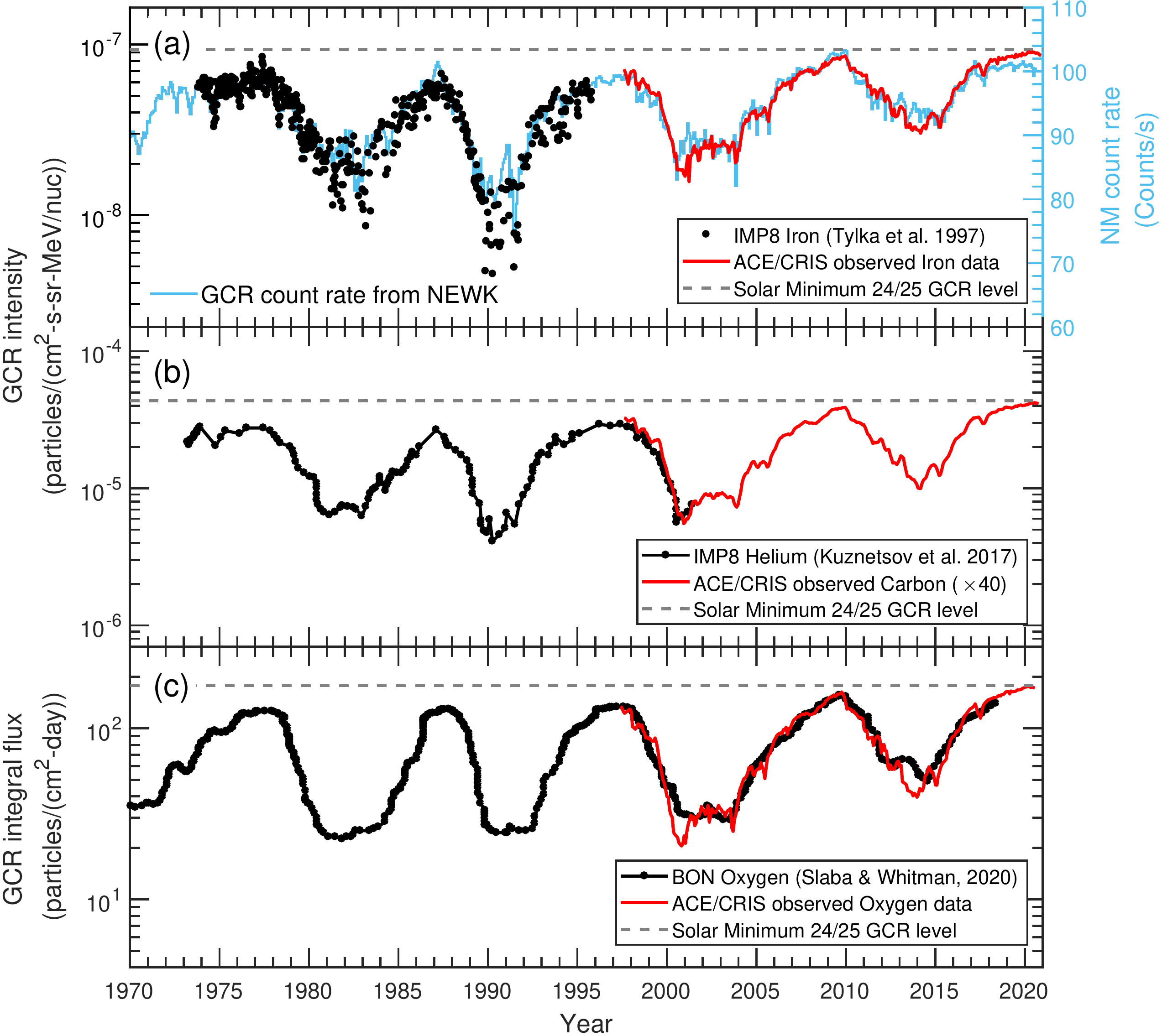}
\caption{Secular variations of GCR intensities in interplanetary space and NM count rates. (a) GCR iron intensities from \emph{ACE}/CRIS at 129.1--428.7 MeV/nuc (\emph{red curve}) and from \emph{IMP}-8 at 97.1--432 MeV/nuc (\emph{black dots}, adapted from \citealt{Tylka97}), compared with Newark GCR count rates (\emph{cyan curve, right axis}). (b) GCR carbon intensities from \emph{ACE}/CRIS at 184.8--200.4 MeV/nuc (\emph{red curve}), compared with GCR helium intensities from \emph{IMP8}/MED at $\sim$265 MeV/nuc (\emph{black curve}, adapted from \citealt{kuznetsov17}). (c) 69.4--237.9 MeV/nuc integral oxygen fluxes from \emph{ACE}/CRIS (\emph{red curve}), compared with the BON2020 modelled fluxes (\emph{black curve}, adapted from \citealt{Slaba+Whitman+2020}). The horizontal dashed line marks the peak GCR intensity in the solar minimum $P_{24/25}$.  \label{Fig3}}
\end{figure}

\subsection{Galactic Cosmic Rays during the Solar Cycle 24}
\label{subsect3.2}

Weak solar activity is expected to reduce GCR modulation level and allows more GCR particles to penetrate the inner heliosphere. Secular GCR observations will be helpful to highlight the unusual heliospheric conditions during the extremely quiet solar minimum $P_{24/25}$. Different from the ground-based NM count rates, the observation of GCRs in space is short and often discontinuous, but those early cosmic-ray records (prior to the launch of \emph{ACE} spacecraft) in interplanetary space can be rebuilt from some other satellites or the modern state-of-the-art numerical models (such as the CR{\`E}ME model and the Badhwar-O'Neill model). Figure \ref{Fig3} shows the extended GCR intensity profiles from 1970 to 2020, in which we compare the iron intensities from \emph{ACE}/CRIS with those from \emph{IMP}-8 in panel (a), the carbon intensities from \emph{ACE}/CRIS with the helium intensities from \emph{IMP}-8 in panel (b), and the oxygen intensities from \emph{ACE}/CRIS with the BON2020 simulations in panel (c). In Figure \ref{Fig3}(a), the GCR count rates from NEWK station are also plotted as a baseline. The peak value of GCR intensities at 1 AU in late 2009 was previously reported to be the highest in the space age \citep{McDonald10,mewaldt10,lave13,leske13}, but it is obvious that the maximum GCR intensities reach new heights during the solar minimum $P_{24/25}$, as depicted by the horizontal dashed line in Figure \ref{Fig3}. The recent anomalous high GCR intensities at 1 AU are essentially a response to the unusual changes in the heliosphere, which will be further discussed in Subsection \ref{subsect3.3}.

\begin{figure}[ht!]
\epsscale{0.9}
\plotone{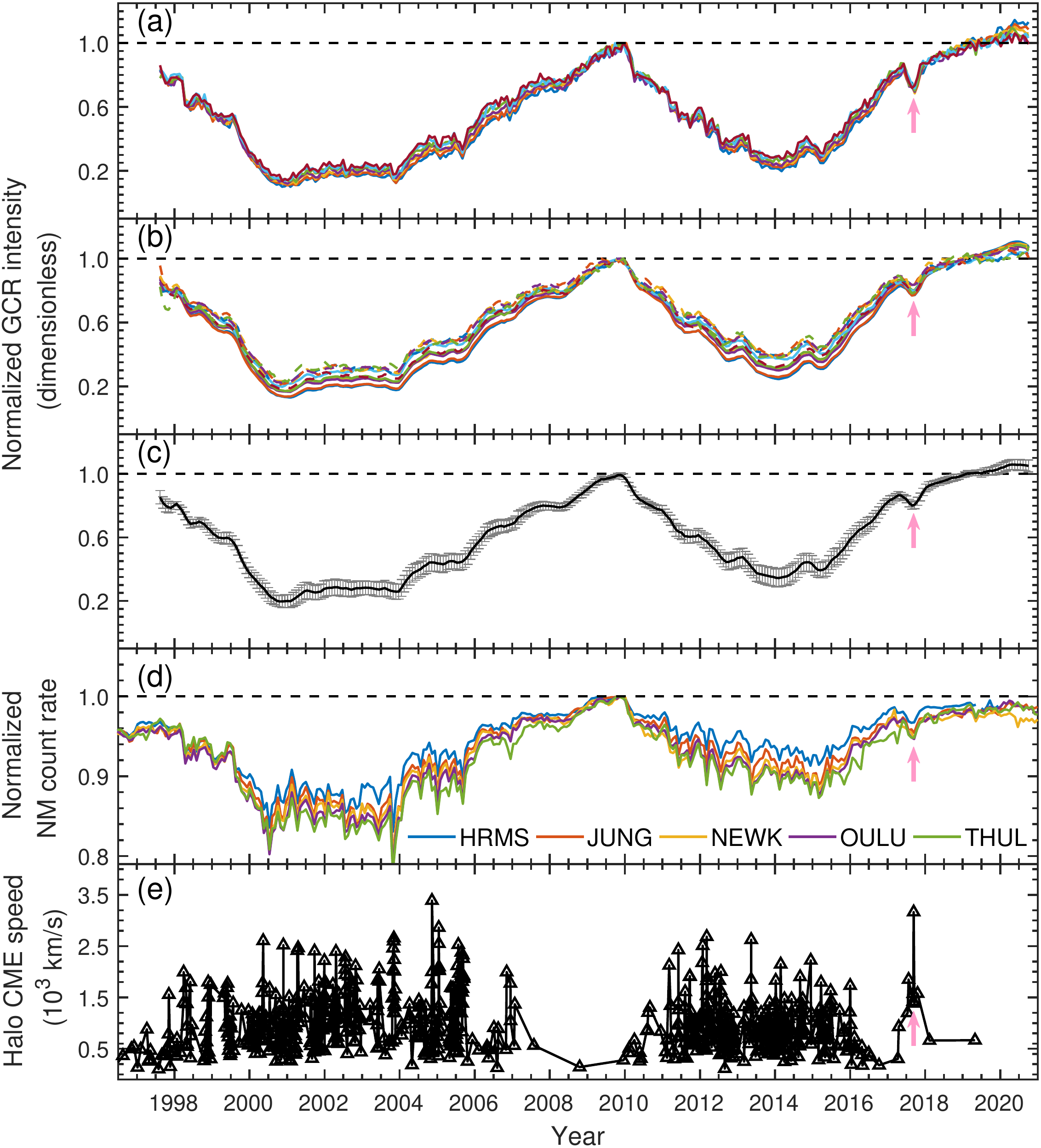}
\caption{(a) \emph{ACE}/CRIS GCR intensities of element oxygen at seven energy intervals (69.4--89.0 MeV/nuc, 91.0--122.5 MeV/nuc, 124.0--150.3 MeV/nuc, 151.6--174.9 MeV/nuc, 176.0--197.3 MeV/nuc, 198.3--218.0 MeV/nuc, 219.1--237.9 MeV/nuc). (b) \emph{ACE}/CRIS GCR intensities of twelve selected GCR species (C, O, Ne, Mg, Si, S, Ar, Ca, Ti, Cr, Fe, Ni). (c) Average intensities of the twelve selected GCR species. (d) Neutron monitor count rates. (e) Historical halo CME events. Note that the intensities in panels (a)--(d) are normalized to their solar minimum $P_{23/24}$ levels, and the horizontal dashed line represents the normalized value equal to 1. The pink-arrow dip in the GCR intensity was caused by CMEs (see Subsection \ref{subsect3.4} for details).  \label{Fig4}}
\end{figure}

Figures \ref{Fig4}(a)--(c) plot the 27-day averaged \emph{ACE}/CRIS GCR intensities over the solar cycles 23 and 24. Panel (a) shows the GCR intensities of element oxygen at seven energy bins, panel (b) the intensities of twelve selected species (C, O, Ne, Mg, Si, S, Ar, Ca, Ti, Cr, Fe, Ni), and panel (c) the averaged intensities of these twelve species. For comparison purpose, all profiles are normalized to the solar minimum $P_{23/24}$. It is clearly noted that the GCR intensities at 1 AU in late 2009 are much higher than those in 1997--1998 but slightly lower than those during the solar minimum $P_{24/25}$. The peak value of GCR intensities in the solar minimum $P_{24/25}$ is $\sim$25\% higher than that in the solar minimum $P_{22/23}$ and $\sim$6\% higher than that in the solar minimum $P_{23/24}$.

Figure \ref{Fig4}(d) plots the 27-day averaged NM count rates from 1997 to 2020. Similar to the GCR intensity profiles, NM count rates also reach the highest level in late 2009 \citep{2010JGRA..11512109M}, revealing that both low ($<$1 GeV) and high energy (several GeV) GCR particles have more chances to reach the Earth during the solar minimum $P_{23/24}$ \citep{mewaldt10}. However, the peak value of NM count rates in 2019--2020 does not exceed its 2009 level, which is not in accord with the observed GCR intensities in interplanetary space. The ground-based NM station only observes the high-energy GCRs (several GeV) as a consequence of the shielding effects from the Earth’s magnetosphere and atmosphere, yet the \emph{ACE}/CRIS instrument measures relatively low-energy particles ($<$1 GeV). The median energy ($E_M$) of the five selected NMs ranges from 10.17 to 12.67 GeV, corresponding to the geomagnetic cutoff rigidity of 0.30 $\sim$ 4.58 GV. The different responses of GCRs in space and at the ground level to varying solar activity may arise from the distinct modulation processes between high and low energy particles as they propagate in the heliosphere. We infer that in the solar minimum $P_{24/25}$, the low-energy GCR particles are more likely to be influenced by the weakening solar modulation compared to the high-energy particles. In addition, the NM count rates are sensitive to the complex changes in the Earth's environmental fluctuations, such as the atmospheric pressure, atmospheric temperature, atmospheric water vapor, instrumental temperature, dynamic magnetospheric condition, and snow effect (see \citealp{2010JGRA..11512109M}, and references therein). The accurate reason for the difference between GCR intensities in interplanetary space and at the ground level is still not known completely. 

Figure \ref{Fig4}(e) plots the historical halo CMEs identified from the Large Angle and Spectrometric Coronagraph (LASCO) on the \emph{Solar and Heliospheric Observatory} (\emph{SOHO}) mission since 1996. We put the discussion of this figure in Subsection \ref{subsect3.4}.

\begin{figure}[ht!]
\epsscale{0.9}
\plotone{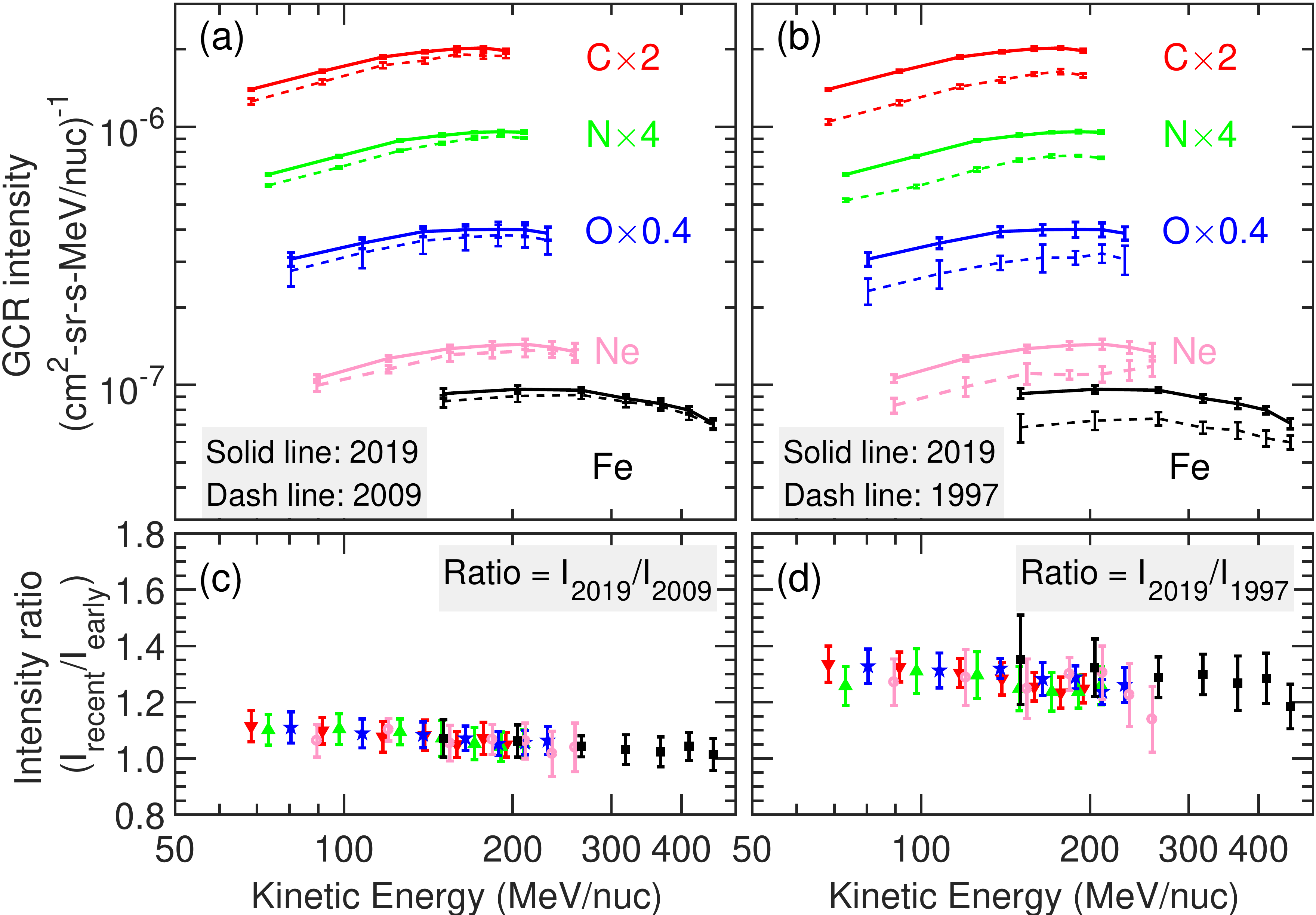}
\caption{(a) and (b) show the comparison of the \emph{ACE}/CRIS GCR differential energy spectra between 2019 and 2009, and between 2019 and 1997, respectively. Energy spectra are multiplied by arbitrary scale factor. Solid and dashed lines are quadratic fits to the experimental data. (c) and (d) show the spectral ratios of the intensity between 2019 and 2009, and between 2019 and 1997, respectively.  \label{Fig5}}
\end{figure}

Figures \ref{Fig5}(a) and \ref{Fig5}(b) display the \emph{ACE}/CRIS observed energy spectra of GCR elements C, N, O, Fe, and Figures \ref{Fig5}(c) and \ref{Fig5}(d) show the spectral ratios of 2019 to 2009, and 2019 to 1997, respectively. During the solar minimum $P_{24/25}$, the significant enhancement in GCR intensity is observed at all lower and higher energies. Although GCR nuclei differ in energy bands, their intensity ratios are relatively consistent. The ratio of 2019 to 2009 is in the range of 1.20 $\sim$ 1.30, and the ratio of 2019 to 1997 is in the range of 1.05 $\sim$ 1.10. This is basically coincident with the previous analysis, in which the peak GCR intensities are greater by $\sim$25\% and $\sim$6\% during the solar minimum $P_{24/25}$ than those in the late 1997 and 2009, respectively. Furthermore, it should be noted that the ratios at lower energies are slightly larger than those at higher energies, indicating that the low-energy GCRs are more sensitive to the variation of solar modulation than the high-energy ones.

\begin{figure}[ht!]
\epsscale{1}
\plotone{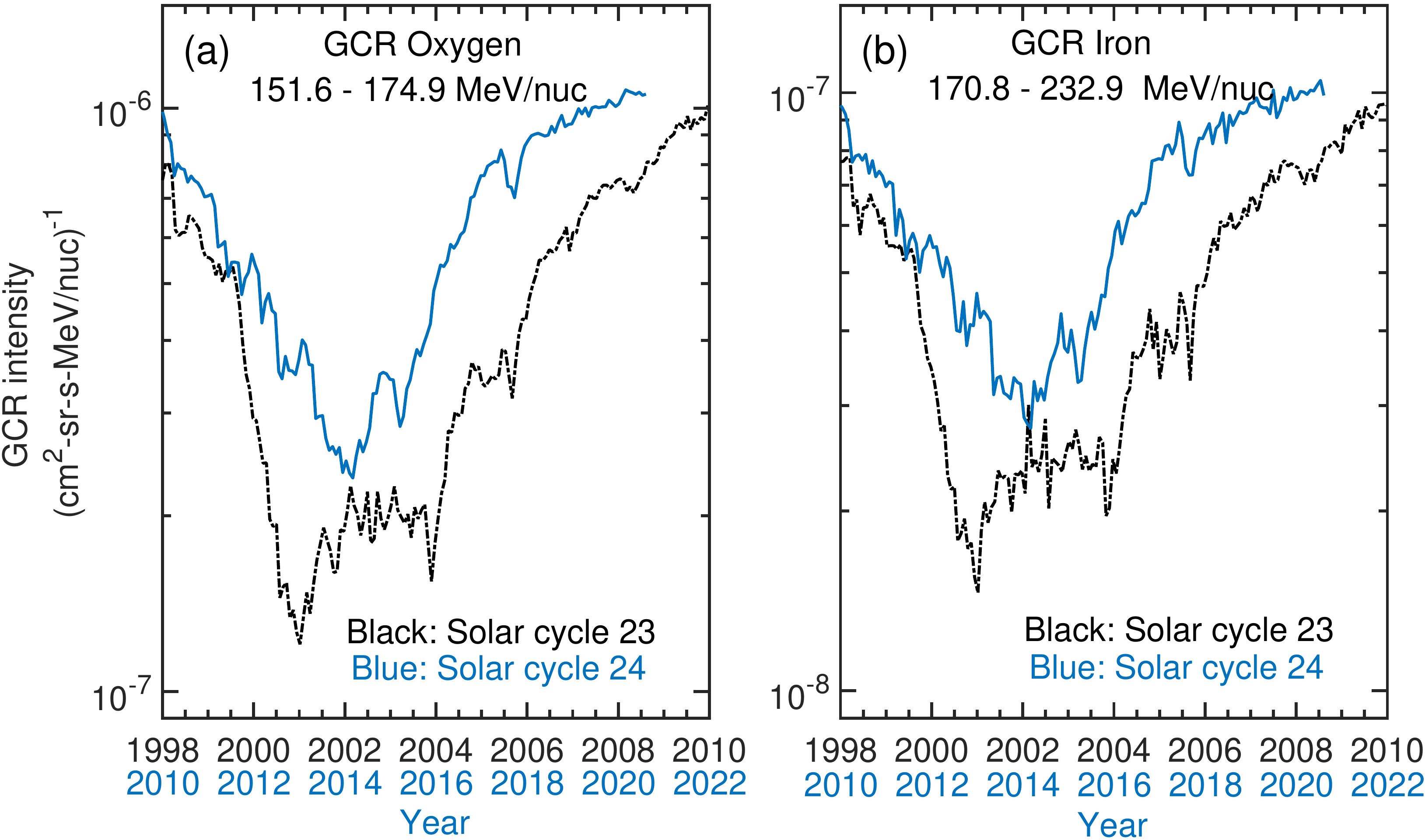}
\caption{Comparison of 27-day averaged GCR intensities at 1 AU between the solar cycles 23 (1998--2010) and 24 (2010--2020), for oxygen (a) and iron (b).  \label{Fig6}}
\end{figure}

In Figure \ref{Fig6}, we compare the \emph{ACE}/CRIS 27-day averaged  GCR intensities between the solar cycles 23 and 24, with panel (a) for 151.6--174.9 MeV/nuc oxygen and panel (b) for 170.8--232.9 MeV/nuc iron. Interestingly, the GCR intensities are significant higher in the solar cycle 24 than in the solar cycle 23, for both ascending and descending phases, which is the desired result of the less active Sun and weakened solar modulation in the solar cycle 24. Furthermore, the bottom of the GCR profile in the solar cycle 23 is deep and wide, but it is shallow and short-lasting in the solar cycle 24. The former reflects a strong and continuous solar modulation, and the latter corresponds to a weakening and short duration solar modulation.

\subsection{Inner Heliospheric Environment during the Solar Cycle 24}
\label{subsect3.3}

The record-breaking GCR intensities at 1 AU in the solar minimum $P_{24/25}$ naturally raise the question of the contributing factors. A general solution to this question is to investigate the inner heliospheric conditions, including the rate of coronal mass ejection (CME), the Sun's polar magnetic field strength, the tilt angle of HCS, the strength and turbulence of the interplanetary magnetic field, as well as the speed and dynamic pressure of the solar wind. 

\subsubsection{Coronal Mass Ejection Rate}
CMEs commonly carry stronger magnetic field than the surrounding solar wind, and cause the violent interplanetary disturbances and magnetic shielding of charged particles when the CME-driven shocks and/or magnetic clouds pass through. Consequently, large CMEs effectively prevent cosmic ray particles from diffusing in the inner heliosphere \citep{wibberenz98,cane00,kilpua17}. Moreover, CMEs are frequently accompanied with other forms of solar activity, such as solar flare \citep{vrsnak16,Syed18}, eruptive prominence and X-ray sigmoids \citep{Gibson02,Pevtsov02}, which further enhances the shielding effectiveness of GCR particles.

\begin{figure}[ht!]
\epsscale{1}
\plotone{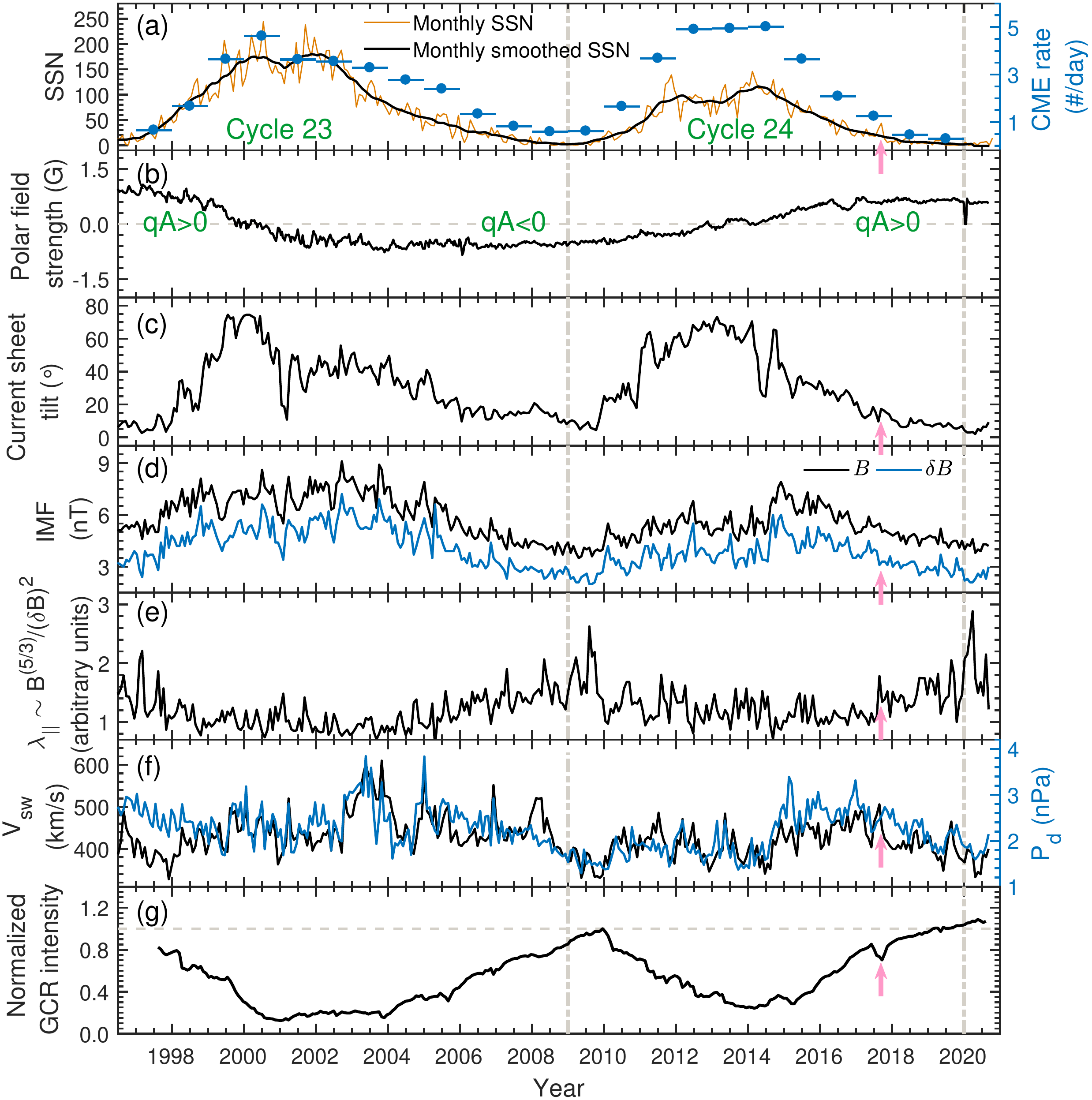}
\caption{27-day averaged solar wind/interplanetary parameters from 1996 to 2020. (a) Sunspot number and annual CME rate. (b) Mean solar polar field strength. (c) HCS Tilt angle. (d) IMF magnitude $B$ and its root mean square $\delta B$. (e) Estimated parallel MFP $\lambda _{\parallel}$. (f) Solar wind speed (\emph{black curve}) and dynamic pressure (\emph{cyan curve}). (g) 69.4--237.9 MeV/nuc GCR oxygen intensity from \emph{ACE}/CRIS (normalized to the solar minimum $P_{23/24}$), and the horizontal dashed line represents the normalized GCR intensity equal to 1. The pink-arrow dip in the GCR intensity was caused by CMEs (see Subsection \ref{subsect3.4} for details).  \label{Fig7}}
\end{figure}

In Figure \ref{Fig7}(a), we present the annual rate of CME acquired from the automated CACTus catalog between 1997 and 2019. It shows that the rate of CME in 2019 is 0.29, which is $\sim$56\% and $\sim$52\% lower than that in 1997 and 2009, respectively. The lower number of CMEs leads to a fewer magnetic irregularities to inhibit GCR diffusion, which in turn contributes to an increase in GCR flux. In addition to the insufficient number, \cite{wang14} suggested that the speed and the mass of the CMEs in the solar cycle 24 are much smaller than those in the solar cycle 23. 
These rare, slow speed and less massive CMEs to some extent weaken the solar modulation of cosmic rays \citep{paouris13}.

\subsubsection{Solar Polar Field}
The polarity of the Sun's polar fields reverses or flips approximately every 11 years during the period of solar maximum \citep{babcock58,babcock59,owens13}. The last two reversals happen around 2001 and 2014, respectively, and it is currently in the positive polarity ($qA>0$). The Sun's polar fields also display anomalous properties in the recent two cycles, for example, the noticeably weakened polar magnetic fields in the solar minimum $P_{23/24}$ \citep{wang09}, the prolonged unusual hemispheric asymmetry of the polar field reversal pattern in the solar cycle 24 \citep{sun15,mordvinov16,janardhan18}. 

The Sun's polar fields influence the penetration of GCRs into the heliosphere by altering the stream structure of the solar wind and the magnitude of the IMF \citep{lee09}. It is reported that the Sun's polar field strength in the solar cycle 23 is $\sim$50\% lower than that in the cycles 21--22, and the flux of the open magnetic field line at the Earth orbit reaches the lowest since 1963 \citep{ahluwalia10}. During the solar cycle 24, the Sun's polar field strength begins to recover after 2014 and enters into a plateau period after the year 2017 with a stable strength $\sim$0.6 G, which is very close to or slightly higher than that in the solar cycle 23, as shown in Figures \ref{Fig2}(b) and \ref{Fig7}(b). It is possible that more GCR particles enter into the heliosphere as a consequence of the weakening solar polar field.

\subsubsection{Tilt Angle of the Helisopheric Current Sheet}
Previous studies have studied the influence of a wavy interplanetary current sheet on the GCR intensity, and found an inverse correlation between the HCS tilt angle and the GCR intensity \citep{jokipii81,potgieter01}. It is because that the GCR particles mainly drift outward ($qA>0$ polarity) or inward ($qA<0$ polarity) along the HCS near the solar equator, and a large tilt angle means that the cosmic ray particles have to drift a longer path length toward the Earth, leading to an increase in the time lag between the source and the observer (\citealp{ferreira04,mewaldt10,zhao14,ross19}). The variation of the HCS tilt angle may have important, even the dominant influences on the GCR modulation process during the solar minimum epoch. In addition, the drift of GCRs along the HCS plays a more significant role than the diffusion when $qA>0$.
 
Figure \ref{Fig7}(c) shows the temporal variation of the HCS tilt angle. The tilt angle begins to decrease after 2014 and reaches the minimum value of $\sim$2.1$^\circ$ in April 2020, which is $\sim$22\% lower than that in the solar minimum $P_{22/23}$ and $\sim$53\% lower than that in the solar minimum $P_{23/24}$. This extremely low tilt angle reflects a very close to the solar equatorial HCS, which results in an enhanced outward drift velocity (for $qA>0$) and the greatly increased GCR intensity. As shown in Figure \ref{Fig7}(g), the HCS tilt angle begin to increase since May 2020 and reaches a value of 14.8$^\circ$ by the end of November 2020, accompanied by the gradual decrease of the GCR flux.

\subsubsection{IMF Strength and Turbulence Level}
The interplanetary magnetic field has important influences on both drifts and the diffusion of the cosmic rays, and the cosmic-ray intensities are found to be negatively correlated with the IMF strength (denoted by $B$). The drift velocity of the GCRs increases with the decrease of $B$ (e.g., \citealp{Jokipii77,Jokipii89}). However, the turbulence in the IMF regulates the pitch-angle scattering of the cosmic rays, where the diffusion coefficient of the cosmic rays is proportional to $1/B$ or some power of $1/B$  (e.g., \citealp{1981ApJ...248.1156J,ferreira04}). The cosmic-ray diffusion associated with the turbulent IMF includes both parallel and perpendicular components, and the relation between diffusion coefficient ($\kappa$) and particle's mean free path ($\lambda$) is given by: 
\begin{equation}
\lambda_{\perp,\parallel} \equiv 3\kappa_{\perp,\parallel} / v,
\label{eq1}
\end{equation}
where $v$ is the particle velocity, $\lambda_{\parallel}$ and $\lambda_{\perp}$ are the parallel and perpendicular MFPs, $\kappa_{\parallel}$ and $\kappa_{\perp}$ are the parallel and perpendicular diffusion coefficients, respectively \citep{zank98,pei10}. The radial MFP $\lambda_{rr}$ governed by the parallel diffusion $\lambda_{\parallel}$ is almost constant within the inner heliosphere \citep{zhao17}. The perpendicular MFP $\lambda_{\perp}$ is nearly three orders of magnitude smaller than the parallel MFP $\lambda_{\parallel}$ \citep{zhao18}. The perpendicular diffusion $\kappa_{\perp}$ is often assumed to scale as the parallel diffusion $\kappa_{\parallel}$ \citep{ferreira04}.

Considering the standard quasi-linear theory (QLT) and assuming magnetostatic turbulence, the parallel MFP $\lambda_{\parallel}$ is proportional to $B^{5/3}/(\delta B)^2$, where $B$ is the mean IMF magnitude and $\delta B$ is the root mean square variation in the vector IMF \citep{zank98,mewaldt10}. With the simplified relation $\lambda _{\parallel} \sim B^{5/3} / (\delta B)^2$, the parallel MFP component $\lambda_{\parallel}$ at 1 AU is estimated over the cycles 23 and 24 (Figure \ref{Fig7}(e)). The IMF magnitude $B$ reaches a minimum value of 3.8 nT in the solar minimum $P_{24/25}$, which is $\sim$24\% smaller than that in the solar minimum of cycles 20 to 22 ($\sim$5 nT) but $\sim$5.6\% larger than that in late 2009 ($\sim$3.6 nT). The IMF turbulence $\delta B$ is $\sim$2.1 nT in the solar minimum $P_{24/25}$, which is $\sim$5\% higher than that in the solar minimum $P_{23/24}$. The weak IMF strength and turbulence lead to a noticeable increase in the estimated cosmic-ray MFP ($\lambda_{\parallel}$), and the maximum value of $\lambda_{\parallel}$ in the solar minimum $P_{24/25}$ exceed those in 1997 by $\sim$31\% and those in 2009 by $\sim$10\%. It is seen that the prominent peak on the $\lambda_{\parallel}$ profile (around March 2020) is followed by the record-breaking GCR intensity. 


\subsubsection{Solar Wind}
The solar wind condition (including speed, density, and temperature) varies over time and  influences the GCR transport throughout the heliosphere. The SW speed affects both the outward convective rate of GCRs and the adiabatic energy loss rate, and generally speaking, a low solar wind speed reduces the outward convection rate and the adiabatic cooling energy loss rate. The GCR intensities are found to be anticorrelated with the solar wind speed, which reveals that the lower the solar wind, the higher the GCR intensities and vice-versa (e.g., \citealp{Ihongo16,zhao14}). 

Figure \ref{Fig7}(f) shows the temporal variation of the SW speed ($V_{sw}$, \emph{black curve}) and the SW dynamic pressure ($P_d$, \emph{blue curve}). The average speeds during the last three solar minima periods are $371 \pm 19$ km/s, $363 \pm 26$ km/s, and $374 \pm 24$ km/s, respectively, and the corresponding dynamic pressures are $2.4 \pm 0.2$ nPa, $1.5 \pm 0.1$ nPa, and $1.8 \pm 0.1$ nPa, respectively. It is clear that in the solar minimum $P_{23/24}$, the SW speed and dynamic pressure reach their lowest values over the past three solar minima, which is thought to be an important reason for the unusual high GCR intensities in late 2009 \citep{leske13,zhao14}. In contrast, the enhanced SW speed and dynamic pressure in the solar minimum $P_{24/25}$ are not conducive to the increase of GCR intensity to a certain extent.
\subsection{A Sudden Dip in the GCR Intensity during the Descending Phase of Cycle 24}
\label{subsect3.4}

During the descending phase of the solar cycle 24, the GCR intensity observed at 1 AU occurs a short-duration but significant dip in the second half of 2017 (around September), after which the GCR intensity continues to rise. This dip is also recorded by the ground-based NM stations, as marked by the pink arrows in Figures \ref{Fig4} and \ref{Fig7}.

The solar wind and interplanetary disturbances are naturally considered to be the causes of this GCR anomaly. The successive CMEs between September 4 to 10, 2017 (including the fastest halo CME on 10 September 2017, see Figure \ref{Fig4}(e)) lead to a highly disturbed interplanetary and geospace environment (\citealp{guo18,lee18,ding20}). Furthermore, the increased sunspot number (Figure \ref{Fig7}(a)), the elevated HCS tilt angle (Figure \ref{Fig7}(c)) and the increased SW speed (Figure \ref{Fig7}(f)) may also result in a drop in the GCR intensity as a consequence of the enhanced solar modulation.

\subsection{GCR Radiation Dose Rate on the Lunar Surface}
\label{subsect3.5}

The deep-space radiation environment, consisting of solar and galactic cosmic rays, poses a great threat to the manned space missions. Therefore, the long-term and the short-term radiation effects must be considered before designing a deep space mission. As a stable background of the energetic particles, GCRs are the most difficult to shield against and can cause severe hazards to astronauts and precision payloads on spacecrafts. \citealp{Schwadron10} previously reported that the GCR dose rates near the lunar surface during the prolonged solar minimum $P_{23/24}$ are the largest since the year 1987. Considering that the peak value of the GCR intensities in the solar minimum $P_{24/25}$ is much higher than that in the 2009 solar minimum, we naturally want to know what the space radiation environment is like now. 

\begin{figure}[ht!]
\epsscale{0.6}
\plotone{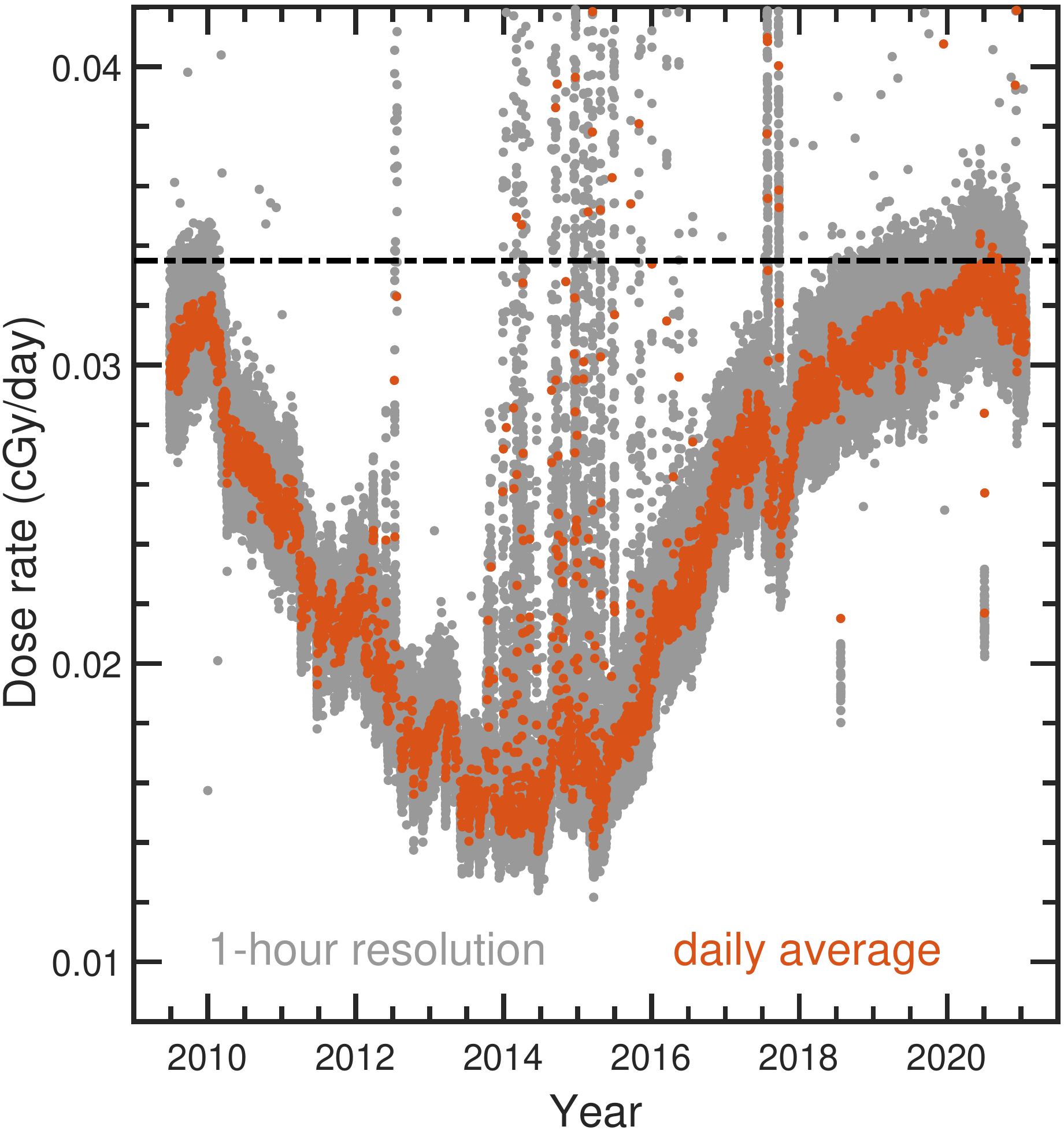}
\caption{GCR radiation dose rates on the lunar surface measured by the \emph{LRO}/CRaTER. The horizontal dash-dotted line marks the peak value of does rate in the solar minimum $P_{24/25}$.  \label{Fig8}}
\end{figure}

The Cosmic Ray Telescope for the Effects of Radiation (CRaTER) instrument on board the \emph{Lunar Reconnaissance Orbiter} (\emph{LRO}) was intended to investigate the lunar radiation environment, and is currently orbiting the Moon at an altitude of 50 km \citep{Spence10}. Since its launch on June 18, 2009, \emph{LRO} has been measuring the lunar radiation environment for nearly 12 years. Figure \ref{Fig8} shows the hourly and daily average dose rates observed by the \emph{LRO}/CRaTER, and the horizontal dash-dotted line is the peak value of dose rates in the solar minimum $P_{24/25}$. It can be seen that the dose rates are tightly correlated with the solar activity as well as the GCR intensities. In the recent solar minimum, the GCR radiation dose rate and the GCR intensity increase substantially because of the weakening solar modulation. We find that the peak value of dose rates in the first half of 2020 is $\sim$5\% higher than that in 2009--2010, which basically corresponds to the $\sim$6\% increase in the GCR intensities measured at 1 AU. We infer that this may be the highest dose rates on the lunar surface since the 1980s, which puts forward higher requirements for radiation shielding and protection in the outer space. Our result can be used to simulate the GCR space radiation risk to human beings in the near-Earth space, and provide a certain reference value in designing the spacecraft materials in future deep space exploration missions.

\section{Conclusions}
\label{sect4}

The Sun was extraordinarily quiet during the recently complete solar cycle 24, including very low sunspot numbers, extremely flat heliospheric current sheet, reduced CME eruption rate, weak interplanetary magnetic field and turbulence level. The observed galactic cosmic ray intensities (both in interplanetary space and at the ground level) are sensitive to the heliospheric conditions, the so-called solar modulation. In this work, we focus on the variations of the GCR intensities for the period from 1997 through 2020 with measurements from the \emph{ACE}/CRIS instrument, and investigate the influence of inner heliospheric environments on the GCR intensities. Besides, the long-term variations of the ground-based NM count rates are also studied. The conclusions are summarized as follows.

(1) The GCR intensities observed at 1 AU reach the highest levels in the solar minimum $P_{24/25}$ since the launch of \emph{ACE} spacecraft, which is $\sim$25\% higher than that in the solar minimum $P_{22/23}$ and $\sim$6\% higher than that in the solar minimum $P_{23/24}$. With the extension of \emph{IMP}-8 measurements and BON2020 numerical simulation results, we find that the GCR intensities during the recent solar minimum are at the highest levels since the space age, having the particle energy from tens to several hundred MeV/nuc. The record-breaking GCR intensities at 1 AU increase the radiation dose rates near the lunar surface by $\sim$5\% in comparison to the solar minimum $P_{23/24}$, from the measurements of \emph{LRO}/CRaTER instrument.

(2) The peak value of NM count rates in the solar minimum $P_{24/25}$ is lower than that in late 2009. The difference between GCR intensities in interplanetary space and NM count rates seems to be relevant to the different modulation processes of the high and low energy particles. We think that the high-energy GCR particles are not heavily modulated as the low-energy ones, which means the low-energy GCR particles are more likely to be influenced by varying solar modulation. An alternate or additional possibility is that the NM count rates are sensitive to the conditions of the Earth's magnetosphere and atmosphere. It remains to be further studied in a future work.

(3) We find that during the solar minimum $P_{24/25}$, the mean solar polar field strength remains weak and is close to the value in the 2009 solar minimum; the HCS tilt angle reaches a minimum value of $\sim$2.1$^{\circ}$ in April 2020, which is $\sim$22\% lower than that in the solar minimum $P_{22/23}$ and $\sim$53\% lower than that in the solar minimum $P_{23/24}$; the CME eruption rate is very small, which is less than half of the CME rate during the solar minima $P_{22/23}$ and $P_{23/24}$; the strength and turbulence of IMF keep at relatively low levels, which causes an $\sim$10\% increase in the estimated cosmic ray MFP ($\lambda_{\parallel}$) compared to that in late 2009. All these factors collectively contribute to less solar modulation and unusual increase in the GCR intensity.

This study is a pure statistical analysis of the GCR variations in the relatively low energy band (50--500 MeV/nuc). In the meanwhile, we discuss the unusual heliospheric conditions in the inner heliosphere in order to explain our findings of the record-breaking GCR intensities at 1 AU. How much of the heliospheric changes influence the GCR intensities needs to be studied. Further work includes the quantitative diagnosis of the effects of drifts, diffusion, and convection on the cosmic-ray modulation from the perspective of numerical simulation (e.g., \citealp{Strauss2014,zhao14,shen18,shen19}).

\acknowledgments

This work is supported by the Science and Technology Development Fund, Macau SAR (File Nos. 008/2017/AFJ, 0042/2018/A2, and 0002/2019/APD) and the National Natural Science Foundation of China (Grant No. 11761161001). 
The authors appreciate beneficial discussions with Dr. Laxman Adhikari at UAH.



\end{document}